\documentclass[submission, Phys]{SciPost}

\hypersetup{
    colorlinks,
    linkcolor={red!50!black},
    citecolor={blue!50!black},
    urlcolor={blue!80!black}
}

\usepackage[bitstream-charter]{mathdesign}
\usepackage{graphbox}
\DeclareSymbolFont{usualmathcal}{OMS}{cmsy}{m}{n}
\DeclareSymbolFontAlphabet{\mathcal}{usualmathcal}

\begin{document}

% TODO: write your article's title here.
% The article title is centered, Large boldface, and should fit in two lines
\begin{center}{\Large \textbf{
Lepton flavor violation with tau leptons\\
}}\end{center}

% TODO: write the author list here. Use initials + surname format.
% Separate subsequent authors by a comma, omit comma at the end of the list.
% Mark the corresponding author with a superscript *.
\begin{center}
Julian Heeck\textsuperscript{$\star$}
\end{center}

% TODO: write all affiliations here.
% Format: institute, city, country
\begin{center}
Department of Physics, University of Virginia,
Charlottesville, Virginia 22904-4714, USA
\\
% TODO: provide email address of corresponding author
* heeck@virginia.edu
\end{center}

\begin{center}
\today
\end{center}

% For convenience during refereeing (optional),
% you can turn on line numbers by uncommenting the next line:
%\linenumbers
% You should run LaTeX twice in order for the line numbers to appear.

\definecolor{palegray}{gray}{0.95}
\begin{center}
\colorbox{palegray}{
  \begin{tabular}{rr}
  \begin{minipage}{0.1\textwidth}
    \includegraphics[width=30mm]{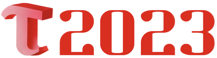}
  \end{minipage}
  &
  \begin{minipage}{0.81\textwidth}
    \begin{center}
    {\it The 17th International Workshop on Tau Lepton Physics}\\
    {\it Louisville, USA, 4-8 December 2023} \\
    \doi{10.21468/SciPostPhysProc.?}\\
    \end{center}
  \end{minipage}
\end{tabular}
}
\end{center}

\section*{Abstract}
{\bf
% TODO: write your abstract here.
%The abstract is in boldface, and should fit in 8 lines.
%It should be written in a clear and accessible style, emphasizing the context, the problem(s) studied, the methods used, the results obtained, the conclusions reached, and the outlook. You can add a table contents, recommended if your paper is more than 6 pages long.\\
We review the status and importance of lepton flavor violation with tauons, focusing on overlooked flavor-breaking patterns as well as tau-flavor violation in nucleon decays.
%well-motivated theoretical models, e.g. connected to neutrino mass, and possible novel signatures involving light new physics.

}

% TODO: include a table of contents (optional)
% Guideline: if your paper is longer that 6 pages, include a TOC
% To remove the TOC, simply cut the following block
%\vspace{10pt}
%\noindent\rule{\textwidth}{1pt}
%\tableofcontents\thispagestyle{fancy}
%\noindent\rule{\textwidth}{1pt}
%\vspace{10pt}

\section{Introduction}
\label{sec:intro}

The successful Standard Model of particle physics features four global $U(1)$ symmetries that were not put in by hand but rather emerge due to the gauge group representations and the restriction to renormalizable interactions:
\begin{align}
U(1)_B \times U(1)_{L_e}\times U(1)_{L_\mu}\times U(1)_{L_\tau} = U(1)_B \times U(1)_{L}\times U(1)_{L_\mu-L_\tau}\times U(1)_{L_\mu+L_\tau - 2 L_e}\,, 
\end{align}
where $L \equiv L_e + L_\mu + L_\tau$ is the total lepton number, $B$ is baryon number, and the individual lepton flavor numbers are denoted by $L_{e,\mu,\tau}$. Charged lepton flavor violation (CLFV) is typically defined as processes involving only charged leptons -- without neutrinos -- that violate the SM symmetry $U(1)_{L_\mu-L_\tau}\times U(1)_{L_\mu+L_\tau - 2 L_e}$~\cite{Heeck:2016xwg}. Even though we know from neutrino oscillations that $U(1)_{L_\mu-L_\tau}\times U(1)_{L_\mu+L_\tau - 2 L_e}$ is broken, this does not lead to any measurable CLFV and hence  leaves $U(1)_{L_\mu-L_\tau}\times U(1)_{L_\mu+L_\tau - 2 L_e}$ as an extremely good approximate symmetry in the charged lepton sector.
Testing this SM prediction would be sufficient motivation by itself, but luckily many SM extensions could actually lead to testable CLFV effects, which elevates CLFV to a very promising handle in the search for new physics~\cite{Davidson:2022jai}.

We will investigate CLFV model-agnostically using higher-dimensional effective operators in the Standard Model Effective Field Theory (SMEFT), which encode the effects of any heavy particles.  CLFV arises first at mass dimension $d=6$ and contains nearly 900 different operators thanks to the three-generational fermion structure of the SM. This is a daunting amount of operators to study and constrain, and of course there are infinitely many more CLFV operators at $d> 6$, although only a small finite number of these lead to observable effects. 

\begin{figure}[h]
\centering
\includegraphics[align=c,width=0.53\textwidth]{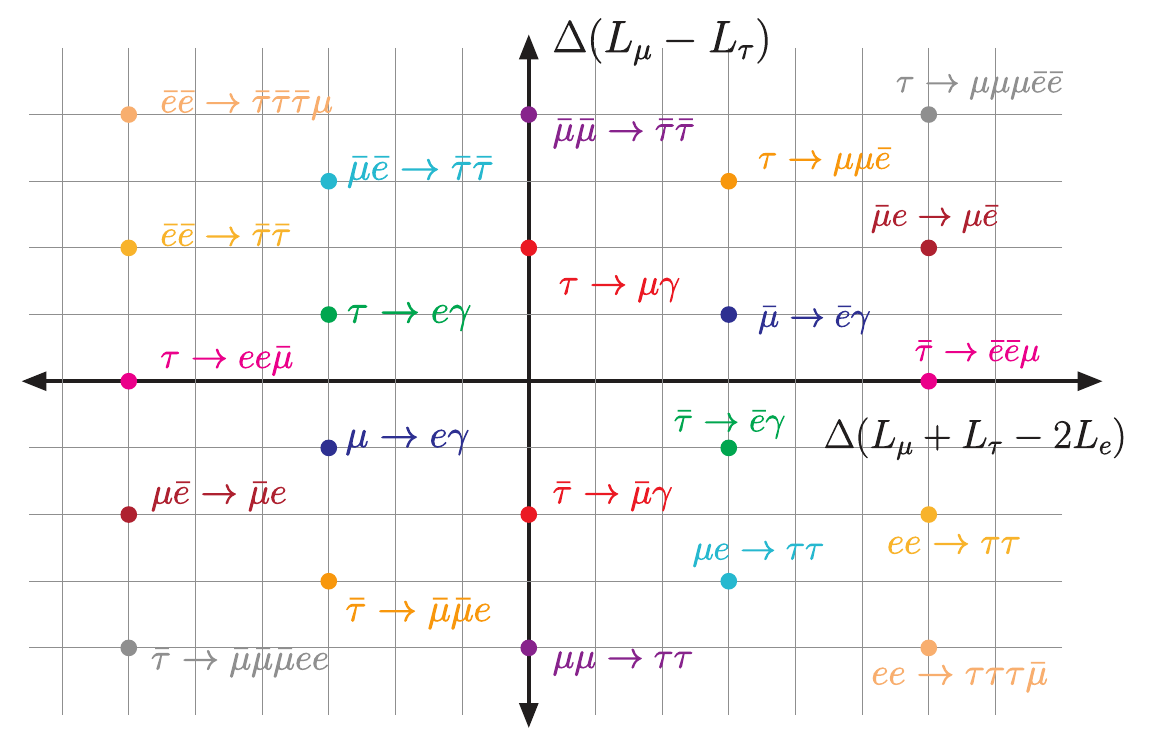}
\includegraphics[align=c,width=0.46\textwidth]{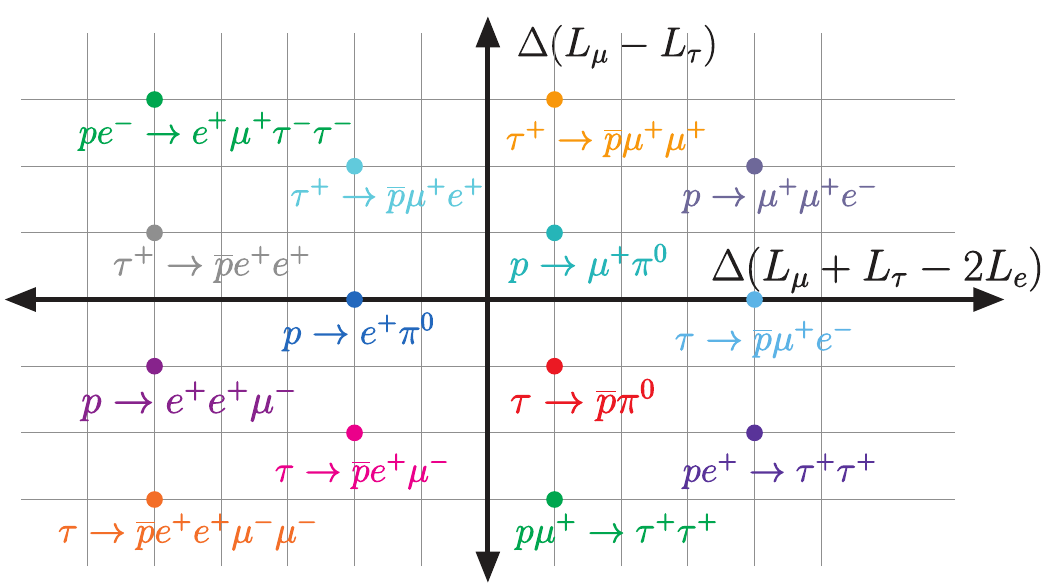}
\caption{Lepton-flavor-violating processes grouped by their $U(1)_{L_\mu-L_\tau}\times U(1)_{L_\mu+L_\tau - 2 L_e}$ breaking, with one example process shown per group. Left: $\Delta B = \Delta L = 0$, adapted from Ref.~\cite{Heeck:2016xwg}.
Right: $\Delta B = \Delta L = 1$, from Ref.~\cite{Heeck:2024jei}.}
\label{grid}
\end{figure}

Complementary to the SMEFT organization of CLFV in terms of mass dimension, we can group all CLFV processes by how they break $U(1)_{L_\mu-L_\tau}\times U(1)_{L_\mu+L_\tau - 2 L_e}$~\cite{Heeck:2016xwg}, resulting in the 2D grid of Fig.~\ref{grid} (left). Here, every dot such as $\mu\to e\gamma$ stands for all processes with the same flavor content, such as $\mu\to e e \bar{e}$, $\mu\to e \gamma\gamma$, $Z\to \bar{\mu}e$, etc, as they can all be obtained by each other by closing loops using SM interactions; which process dominates in each group depends on the underlying model. Also note that the processes involving two leptons (the six groups closest to the origin) and four leptons (the 12 groups around those first six) all arise already at $d=6$, whereas anything beyond requires $d\geq 10$ and is hence far more suppressed.

Since we do not know if and how $U(1)_{L_\mu-L_\tau}\times U(1)_{L_\mu+L_\tau - 2 L_e}$ is broken in the charged lepton sector, it is imperative to probe CLFV in as many directions as possible. It is worth pointing out that one can easily impose a flavor symmetry, for example lepton triality~\cite{Bigaran:2022giz} or a discrete or continuous subgroup of $U(1)_{L_\mu-L_\tau}\times U(1)_{L_\mu+L_\tau - 2 L_e}$, to forbid many or even all but one group in Fig.~\ref{grid} (left), so CLFV could well be hiding in groups not under experimental scrutiny. This is important to emphasize because otherwise one could simply put all eggs in the $\mu\to e\gamma$ basket, which is by far the most sensitive CLFV group. Alas, studying $\mu\to e\gamma$ can only tell us if $L_\mu$ and $L_e$ are broken by one unit. CLFV in the tauon sector probes many more breaking patterns and is perfectly \emph{complementary} to muon CLFV.
The previous $B$ factories BaBar and Belle continue to analyze their collected data and improve their limits~\cite{Belle:2023ziz}, while Belle II has now collected enough data to confirm and improve on those constraints~\cite{Banerjee:2022vdd} (see talk by Alberto Martini); LHC experiments have also reached competitive limits the CLFV tauon channel $\tau\to 3\mu$.

\section{CLFV by two units}

The groups closest to the origin in Fig.~\ref{grid} (left) have received the lion share of experimental and theoretical attention and are well covered in other reviews~\cite{Davidson:2022jai}; let us instead focus on the twelve groups \emph{around} them, which all violate at least one lepton flavor by \emph{two} units~\cite{Heeck:2024uiz}. These also arise at $d=6$ in the SMEFT and are thus equally important, but require -- in part -- drastically different search strategies. The exception to this statement are the processes $\tau\to ee \bar\mu$ and $\tau\to \mu\mu\bar e$, which are already covered in tauon CLFV at $B$ factories~\cite{Banerjee:2022vdd}.
Another familiar entry is $\bar\mu e\to \mu \bar e$, which induces muonium--antimuonium conversion and can be searched for with increased precision at the future MACE~\cite{Bai:2022sxq} facility. Notice, however, that this conversion only efficiently probes two of the three $d=6$ SMEFT operators with $\Delta L_\mu = -\Delta L_e = 2$. The remaining blind direction only yields heavily suppressed muonium conversion rates~\cite{Conlin:2020veq} and is better constrained by searches for lepton flavor universality, e.g.~$\Gamma (\mu\to e \nu\nu)/\Gamma(\tau \to e \nu\nu)$. Even though these observables involve neutrinos, they can put the strongest constraint on some CLFV operators~\cite{Heeck:2024uiz} and can be further improved with Belle-II data on $\tau\to \ell\nu\nu$.

This only leaves CLFV operators with $\Delta L_\tau = 2$, which are considerably more difficult to constrain because they do not induce any tree-level on-shell lepton decays, and the study of ``tauonium'' is impossible.
For all $d=6$ operators involving \emph{left-handed} tauons, one can again obtain reasonable constraints from lepton universality ratios such as $\Gamma (\tau\to e \nu\nu)/\Gamma(\tau \to \mu \nu\nu)$, assuming one operator at a time~\cite{Heeck:2024uiz}. 
The only $\Delta L_\tau = 2$ operators that cannot currently be constrained in a way compatible with the SMEFT assumption, i.e.~to Wilson coefficients below $(100\text{\,GeV})^{-2}$, are the three operators involving \emph{right-handed} tauons~\cite{Heeck:2024uiz}: $\bar{\ell}_\alpha \gamma^\sigma P_R \ell_{\tau}\, \bar{\ell}_\beta \gamma_\sigma P_R  \ell_{\tau} $, with $\alpha,\beta\in \{e,\mu\}$.
Closing loops to map these back onto left-handed operators comes with a strong suppression by lepton Yukawas; letting one tauon be off-shell brings in an expensive $G_F$ suppression. Either way, the limits on these Wilson coefficients are irrelevantly weak. Slightly better are lepton-flavor-violating $Z$ decays $Z\to \tau^+\tau^+ \ell_\alpha^- \ell_\beta^-$, which have never been searched for but could eventually yield SMEFT-relevant limits at a future $Z$ factory. Until then, UV realizations of these operators can still lead to testable signals as long as the mediator particles are below the electroweak scale~\cite{Altmannshofer:2016brv,Bigaran:2022giz}, so experimental searches should commence.

\section{Baryon number violation involving tauons}

So far we have restricted ourselves to the study of CLFV, which in particular assumed processes that conserve baryon and total lepton number. Let us loosen this restriction and consider processes that violate baryon number $B$ by one unit. This leads to proton decay, subject to unfathomably strong limits due to the relative ease with which we can observe an extremely large number of protons for long periods of time~\cite{FileviezPerez:2022ypk}. Angular momentum conservation requires protons to decay into an odd number of leptons, kinematically restricted to electrons, muons, and neutrinos. Once again, SMEFT operators inducing such decays already appear at $d=6$ and take the form $qqq \ell$. In particular, each operator inevitably violates both lepton number and lepton\emph{ flavor}. Even though this is rarely emphasized, it allows us to translate our previous discussion -- and  Fig.~\ref{grid} -- to the baryon-number violating case, and organize $\Delta B$ processes by their flavor content~\cite{Hambye:2017qix}, see Fig.~\ref{grid} (right). The same logic applies: every dot on the grid stands for an entire group and every group probes a different flavor breaking pattern, making it crucial to investigate in all possible direction to not miss new physics.

Just like in the CLFV case, tauon modes are crucial to explore the entire flavor landscape. Tau decays such as $\tau\to \bar{p}\pi^0$ or $\tau\to \bar{p} e^+\mu^-$~\cite{Belle:2020lfn} are clean neutrinoless probes of the underlying operators. However, as pointed out long ago by Marciano~\cite{Marciano:1994bg}, the underlying operators or UV-complete models inevitably also induce nucleon decays into tau neutrinos, which provide far better limits.
In Ref.~\cite{Heeck:2024jei} we have quantitatively explored and confirmed this argument. For operators involving left-handed tauons, two-body $\tau$ decays compete directly with two-body neutron decays such as $n\to \pi^0\bar{\nu}_\tau$, extremely well constrained by Super-K~\cite{Super-Kamiokande:2013rwg}. However, for \emph{right-handed} tauon operators it is possible to suppress nucleon decays quite a bit, forcing them into the untested three-body channel $p\to \eta \pi^+ \bar{\nu}_\tau$~\cite{Heeck:2024jei}. Still, old inclusive nucleon-decay limits~\cite{Heeck:2019kgr} far exceed $\tau\to \bar{p}\eta$ sensitivities.

At $d=7$ in the SMEFT, one can find operators involving two strange quarks that induce the relatively clean two-body decays $\tau\to \Xi \pi$, with nucleon decays forced into the heavily suppressed five-body decay $p\to K^+\mu^+\nu_\mu \pi^-\nu_\tau$. No studies for either of these decays exist, but tauon decays could come close to nucleon-decay sensitivity, so we strongly encourage searches.

\section{Conclusion}

Charged lepton flavor violation probes an important prediction of the SM and arises generically in many SM extensions at testable rates. While CLFV in the muon sector is particularly clean and capable of reaching very high scales, we emphasized here that CLFV in the tauon sector is perfectly complementary and able to probe many more patterns of flavor symmetry breaking, also in the baryon-number-violating case. We proposed a variety of useful searches at Super-K and Belle II to probe overlooked channels.

\section*{Acknowledgements}

I thank Mikheil Sokhashvili and Dima Watkins for collaboration on some of the work presented here.
% and the Tau2023 organizers for a fantastic workshop.
This work was supported in part by the National Science Foundation under Grant PHY-2210428 and a 4-VA at UVA Collaborative Research Grant.

% TODO: include author contributions
%\paragraph{Author contributions}
%This is optional. If desired, contributions should be succinctly described in a single short paragraph, using author initials.

% TODO: include funding information
%\paragraph{Funding information}
%Authors are required to provide funding information, including relevant agencies and grant numbers with linked author's initials. Correctly-provided data will be linked to funders listed in the \href{https://www.crossref.org/services/funder-registry/}{Fundref registry}.

% TODO:
% Provide your bibliography here. You have two options:

% FIRST OPTION - write your entries here directly, following the example below, including Author(s), Title, Journal Ref. with year in parentheses at the end, followed by the DOI number.
%\begin{thebibliography}{99}
%\bibitem{1931_Bethe_ZP_71} H. A. Bethe, {\it Zur Theorie der Metalle. i. Eigenwerte und Eigenfunktionen der linearen Atomkette}, Zeit. f{\"u}r Phys. {\bf 71}, 205 (1931), \doi{10.1007\%2FBF01341708}.
%\bibitem{arXiv:1108.2700} P. Ginsparg, {\it It was twenty years ago today... }, \url{http://arxiv.org/abs/1108.2700}.
%\end{thebibliography}

% SECOND OPTION:
% Use your bibtex library
% \bibliographystyle{SciPost_bibstyle} % Include this style file here only if you are not using our template
\bibliography{SciPost_BiBTeX_File.bib}

\nolinenumbers

\end{document}